\documentclass{aa}
\usepackage{graphicx}
\usepackage{ulem}

      \def\new#1 {{\bf #1 }}
      \def\cut#1 {\sout{#1} }

\begin{document}
\def\ffam {\hbox{$\,.\!\!^{\prime}$}}
\def\ffas {\hbox{$\,.\!\!^{\prime\prime}$}}
\def\ffM {\hbox{$\,.\!\!^{\rm M}$}}
\def\ffm {\hbox{$\,.\!\!^{\rm m}$}}
\def\ffs {\hbox{$\,.\!\!^{\rm s}$}}

\title{The kinetic temperature of a molecular cloud at redshift 0.7:
       Ammonia in the gravitational lens B0218+357
       \thanks{Based on observations with the 100-m telescope of the
       Max-Planck-Institut f{\"u}r Radioastronomie (MPIfR) at Effelsberg}}


\author{C.~Henkel\inst{1}
        \and
        N.~Jethava\inst{1}
        \and
        A.~Kraus\inst{1}
        \and
        K.M.~Menten\inst{1}
        \and
        C.L.~Carilli\inst{2}
        \and
        M.~Grasshoff\inst{3}
        \and
        D.~Lubowich\inst{4}
        \and
        M.J.~Reid\inst{5}}

\offprints{C. Henkel,
\email{chenkel@mpifr-bonn.mpg.de}}

\institute{Max-Planck-Institut f\"ur Radioastronomie, Auf dem H\"ugel 69, D-53121 Bonn, Germany
           \and
           National Radio Astronomy Observatory, Socorro, New Mexico
87801, USA
           \and
           Visiting the Max-Planck-Institut f\"ur Radioastronomie, Auf dem H\"ugel 69, D-53121 Bonn, Germany
           \and
           Department of Physics and Astronomy, Hofstra University, Hempstead, NY\,11549, USA  
           \and
           Harvard-Smithsonian Center for Astrophysics, 60 Garden St., MS 42, Cambridge, MA\,02138, USA}

\date{Received date / Accepted date}

\abstract{Using the Effelsberg 100-m telescope, absorption in the ($J$,$K$) = (1,1), (2,2) and (3,3) inversion lines of ammonia 
(NH$_3$) was detected at a redshift of $z$ = 0.6847 toward the gravitational lens system B0218+357. The $\lambda$$\sim$2\,cm 
absorption peaks at 0.5--1.0\% of the continuum level and appears to cover a smaller fraction of the radio continuum background 
than lines at millimeter wavelengths. Measured intensities are consistent with a rotation temperature of $\sim$35\,K, corresponding 
to a kinetic temperature of $\sim$55\,K. The column density toward the core of image A then becomes $N$(NH$_3$) $\sim$ 
1$\times$10$^{14}$\,cm$^{-2}$ and fractional abundance and gas density are of order $X$(NH$_3$) $\sim$ 10$^{-8}$ and $n$(H$_2$) $\sim$ 
5$\times$10$^{3}$\,cm$^{-3}$, respectively. Upper limits are reported for the (2,1) and (4,4) lines of NH$_3$ and for transitions 
of the SO, DCN, OCS, SiO, C$_3$N, H$_2$CO, SiC$_2$, HC$_3$N, HC$_5$N, and CH$_3$OH molecules. These limits and the kinetic 
temperature indicate that the absorption lines are not arising from a cold dark cloud but from a warm, diffuse, predominantly 
molecular medium. The physical parameters of the absorbing molecular complex, seen at a projected distance of $\sim$2\,kpc to the 
center of the lensing galaxy, are quite peculiar when compared with the properties of clouds in the Galaxy or in nearby 
extragalactic systems. 
}

\titlerunning{Molecular absorption lines in B0218+357}

\authorrunning{Henkel et al.}

\maketitle


\section{Introduction}

With an image separation of 334\,mas  B0218+357 is one of the most compact gravitational lens systems known to date (O'Dea et al. 
\cite{odea92}; Patnaik et al. \cite{patnaik93}). The lensed source, possibly a BL Lac object (Kemball et al. \cite{kemball01}), 
is located at a redshift of $z$ $\sim$ 0.94 (Cohen et al. \cite{cohen03}) and shows a complex radio structure with two dominating 
compact sources, A and B, and an Einstein ring (e.g. Biggs et al. \cite{biggs01a}). A time delay of order 10 days has been measured 
between the two compact sources (Corbett et al. \cite{corbett96}; Biggs et al. \cite{biggs99}, \cite{biggs01b}; Cohen et al. 
\cite{cohen00}). Absorption from the lensing galaxy, a face-on spiral likely of type Sa/Sab (York et al. 2005), is observed 
at a redshift of $z$=0.68466 at optical wavelengths (Brown et al. \cite{browne93}; Stickel \& K{\"u}hr \cite{stickel93}) 
and, remarkably, in the $\lambda$21\,cm line of H{\sc i} (Carilli et al. \cite{carilli93}) and in numerous radio- and 
millimeter-wavelength lines from a variety of molecules. The absorption is observed against the compact component A which is, at 
radio waves, $\sim$3 times stronger than component B (e.g. Patnaik et al. \cite{patnaik93}). Detected molecular species are CO, 
HCN, and HCO$^{+}$ (Wiklind \& Combes \cite{wiklind95}), H$_2$CO (Menten \& Reid \cite{menten96}), H$_2$O (Combes \& Wiklind 
\cite{combes97}), CS (Combes et al. \cite{combesetal97}), OH (Kanekar et al. \cite{kanekar03}) and tentatively also LiH (Combes 
\& Wiklind \cite{combes98}). The molecular spectra, combined with 21\,cm H{\sc i} absorption profiles, have been used to constrain 
the temporal evolution of the fine structure constant (e.g. Carilli et al. \cite{carilli00}; Murphy et al. \cite{murphy01}; 
Kanekar \& Chengalur \cite{kanekar04}). The nature of the cloud, however, is poorly understood.

One of the less well known parameters of extragalactic molecular clouds is the kinetic gas temperature. With area filling factors 
that are not well known, a thermalized tracer like CO cannot be used to determine $T_{\rm kin}$. Better thermometers are symmetric 
top molecules, where relative level populations are determined predominantly by collisions. The two most prominent such molecules 
are NH$_3$ and CH$_3$CN. While extragalactic CH$_3$CN was so far only detected in NGC\,253 (Mauersberger et al. \cite{mauers91}), 
various NH$_3$ `inversion' lines, from the ($J$,$K$) = (1,1) up to the (6,6) and even the (9,9) line, have now been observed in the 
nuclear regions of nearby ($z$$\la$0.001) galaxies (e.g. Martin \& Ho \cite{martin86}; Henkel et al. \cite{henkel00}; Takano et al. 
\cite{takano00}, \cite{takano02}; Wei{\ss} et al. \cite{weiss01}; Mauersberger et al. \cite{mauers03}).

B0218+357 provides a unique view onto a molecular cloud seen at about half a Hubble time in the past ($\Lambda$-cosmology with
$H_{0}$=71\,km\,s$^{-1}$\,Mpc$^{-1}$, $\Omega_{\rm m}$ = 0.27 and $\Omega_{\Lambda}$ = 0.73; Spergel et al. 2003). In an attempt 
to constrain the physical properties of this cloud at a luminosity distance of $\sim$3\,Gpc, we have searched for NH$_3$ and 
other molecular species.

\begin{table}
\caption[]{Observational parameters}
\begin{flushleft}
\begin{tabular}{cccc}
\hline
$\nu$       &  $\theta_{\rm b}^{\rm a)}$ & $T_{\rm sys}^{\rm b)}$ & $\eta_{\rm a}^{\rm c)}$ \\
(GHz)       &  (arcsec)                  &         (K)            &                         \\
\hline
            &                            &                        &                         \\
   5.9      &   130                      &  35                    & 0.53                    \\
13.7--17.8  &   65--50                   &  50                    & 0.40                    \\
21.6--25.8  &   40--35                   &  50                    & 0.35                    \\
43.0--43.2  &   22                       & 125                    & 0.16                    \\
            &                            &                        &                         \\

\hline
            &                            &                        &                         \\
\end{tabular}
\begin{scriptsize}
\item[a)] Full width to half power (FWHP) beam widths  
\item[b)] System temperatures in units of antenna temperature ($T_{\rm A}^{*}$)
\item[c)] Aperture efficiencies, see the Effelsberg calibration pages in {\it www.mpifr-bonn.mpg.de}    
\end{scriptsize}
\end{flushleft}
\end{table}

\section{Observations}

All observations were made with the 100-m telescope at Effelsberg/Germany. In August 2001 and June 2002 we searched, employing 
a single channel $\lambda$$\sim$1.9\,cm HEMT receiver, for ammonia (NH$_3$), sulfur monoxide (SO), methanol (CH$_3$OH), 
the SiC$_2$ radical, cyanodiacetylene (HC$_5$N) and carbonylsulfide (OCS). The measurements were carried out in a position 
switching mode. In December 2001 we used a dual channel 1.3\,cm HEMT receiver to search for cyanoacetylene (HC$_3$N) and silicon 
monoxide (SiO) in its ground vibrational state (see Tables 1 and 2). These measurements were made in a dual beam switching mode 
with a beam throw of 2$'$ and a switching frequency of $\sim$1\,Hz. In January 2002, we also employed a single-channel 7\,mm HEMT 
receiver to search for formaldehyde (H$_2$CO) and deuterated hydrogen cyanide (DCN). These observations were made in a position 
switching mode. Most recently, in June 2003, a two channel 5\,cm HEMT receiver was used to search for the $N$=1--0 $J$=3/2--1/2 
transition of the C$_3$N radical.  

Frequencies, beamwidths, system temperatures and aperture efficiencies are given in Table 1. For all measurements we employed an 
`AK\,90' autocorrelator with eight spectrometers, using bandwidths of 40\,MHz and 512 channels (20\,MHz and 1024 channels 
and 80\,MHz and 256 channels at 5\,cm and 7\,mm, respectively). Calibration was obtained from measurements of NGC~7027 (Ott et al. 
\cite{ott94}). Pointing corrections could be obtained toward B0218+357 itself and were accurate to better than 10$''$.

\section{Results}

Fig.\,1 shows the measured NH$_3$ profiles. The detected inversion lines, with the ordinate displaying absorption in 
units of 1\% of the continuum flux, are characterized by a prominent narrow component with a width of a few km\,s$^{-1}$, 
centered at slightly positive velocities, and a wider and weaker component that is centered at slightly negative velocities 
with respect to a redshift of $z$=0.68466. Tables 2 and 3 display noise levels for all measured spectra and line parameters 
for the detected transitions. The weak but wide velocity component has a 2--3 times higher integrated intensity than the more 
prominent narrow one. At its peak, the narrow component reaches 0.5--1.0\% of the total continuum flux density, while the 
broad component peaks at $\sim$0.3\% in the (1,1) and (2,2) lines and at an even lower level in the (3,3) transition. The
narrow component must be optically thin; otherwise the hyperfine satellites would be visible (for the (1,1) line at $\pm$8 
and $\pm$19\,km\,s$^{-1}$ with respect to the main feature). Due to a lower signal-to-noise ratio, the optical depth of the wide 
component is poorly constrained (see also Sect.\,4.1).  

At $\lambda$=1.9 and 1.3\,cm, continuum flux densities were 1.0--1.2\,Jy, in good agreement with Patnaik et al. \cite{patnaik93}
and Menten \& Reid (\cite{menten96}). At 6\,GHz we find 1.45\,Jy with an error of $\pm$10\%, while no flux density was determined 
at 7\,mm. 

Fig.\,2 shows a Boltzmann plot (rotation diagram) including the four measured `metastable' ($J$=$K$) inversion lines of 
ammonia. Applying
$$
N(J,K)/T_{\rm ex} = 1.61\times 10^{14}\ \frac{J(J+1)}{K^2 \nu} \int{\tau {\rm d}V}
$$
($N(J,K)$ in cm$^{-2}$, $T_{\rm ex}$ in K, $\nu$ in GHz; see e.g. H{\"u}ttemeister et al. 1995) and assuming optically thin
absorption (see Sect.\,4.1) and equal excitation temperatures ($T_{\rm ex}$) across the measured inversion doublets, rotation 
temperatures are of order 35--40\,K. Although the (1,1) and (2,2) lines belong to the para- and the (3,3) line to the ortho-species
(the conversion from one to the other species takes very long; see Cheung et al. 1969), no difference in excitation is apparent. 
The upper limits to the (4,4) line are consistent with the rotation temperature derived from the lower inversion transitions.

\begin{figure}[t]
\vspace{-0.3cm} \centering
\resizebox{22.0cm}{!}{\rotatebox[origin=br]{-90}{\includegraphics{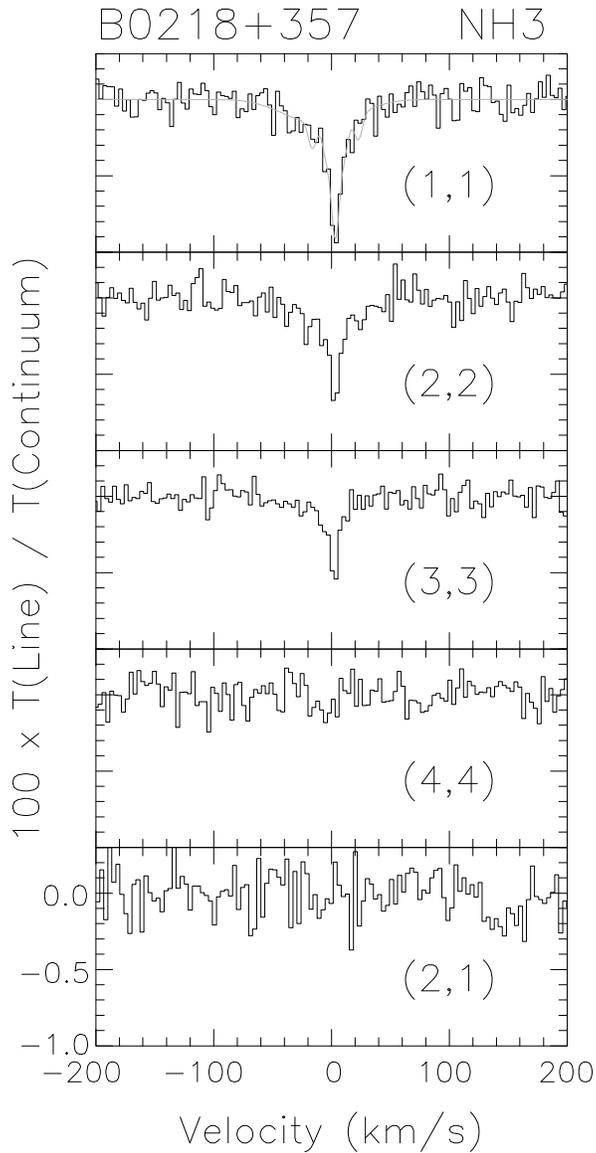}}}
\vspace{-0.1cm}
\caption{NH$_3$ lines with a velocity scale relative to $z$=0.68466, observed towards B0218+357. Channel widths are 3.3\,km\,s$^{-1}$. 
The ordinate is in units of the radio continuum flux. In the case of the (1,1) line a two component fit has been added accounting 
for NH$_3$ hyperfine splitting (Kukolich \cite{kukolich67}). The fit demonstrates that the line profile is consistent with two optically 
thin velocity features. Limited signal-to-noise ratios do not constrain the fit sufficiently to justify an inclusion of its parameters 
in Table 3.
\label{fig1}}
\end{figure}

\begin{table}
\caption{rms line to continuum ratios in units of 10$^{3}$$\times$[rms/$T_{\rm c}$] (Col.\,4) and corresponding channel spacings 
(Col.\,5)}
\begin{center}
\begin{tabular}{ccrrc}
\hline
Molecule & Line                            &\multicolumn{1}{c}{$\nu$} &\multicolumn{1}{c}{rms}& Channel        \\
         &                                 &                          &                       &  width         \\
         &                                 &         (GHz)            & (10$^{-3}$)           & (km\,s$^{-1}$) \\
\hline
C$_3$N   & $N$=1--0, $J$=3/2--1/2          &         5.868                   &     3.0     & 2.29           \\
NH$_3$   & ($J,K$)=(2,1)                   &        13.711                   &     2.1     & 1.71           \\
CH$_3$OH & $J_{\rm K}$=9$_2-10_1$ A$^+$    &        13.724                   &     1.8     & 1.71           \\
SiC$_2$  & $J_{\rm KaKc}$=1$_{01}-0_{00}$  &        14.009                   &     2.0     & 1.65           \\
NH$_3$   & ($J,K)$=(1,1)                   &        14.065                   &     1.0     & 1.65           \\
NH$_3$   & ($J,K)$=(2,2)                   &        14.082                   &     1.1     & 1.65           \\
NH$_3$   & ($J,K)$=(3,3)                   &        14.169                   &     0.7     & 3.30           \\
HC$_5$N  & $J$=9--8                        &        14.225                   &     1.3     & 3.30           \\
NH$_3$   & ($J,K)$=(4,4)                   &        14.329                   &     1.4     & 1.65           \\
OCS      & $J$=2--1                        &        14.440                   &     1.9     & 1.65           \\
SO       & $J_{\rm K}$=1$_0-0_1$           &        17.809                   &     2.6     & 1.32           \\
HC$_3$N  & $J$=4--3                        &        21.602                   &     8.2     & 1.08           \\
SiO      & $J$=1--0, $v$=0                 &        25.776                   &    20.2     & 0.91           \\
DCN      & $J$=1--0                        &        42.985                   &    16.7     & 2.18           \\
H$_2$CO  & $J_{\rm KaKc}$=1$_{01}-0_{00}$  &        43.236                   &   222.7     & 2.17           \\
\hline
\end{tabular}
\end{center}
\end{table}

\begin{table}
\caption{NH$_3$ line parameters$^{\rm a)}$}
\begin{center}
\begin{tabular}{c c r r @{\ } r}
\hline
Line         & $\int{\tau {\rm d}V}$ & $V$\ \ \ \ \ \ \  & \multicolumn{2}{c}{$\Delta V_{1/2}$} \\
             & \multicolumn{4}{c}{(km\,s$^{-1}$)}                        \\
\hline
NH$_3$ (1,1) & --0.062\,(0.013)         &  +3.3\,(0.5) &  9.3&(1.8)        \\
             & --0.183\,(0.019)         & --2.0\,(2.7) & 56.6&(7.8)        \\
NH$_3$ (2,2) & --0.037\,(0.009)         &  +3.1\,(0.5) &  6.6&(1.6)        \\
             & --0.143\,(0.017)         & --1.7\,(3.0) & 54.7&(7.6)        \\
NH$_3$ (3,3) & --0.034\,(0.011)         &  +2.4\,(0.5) &  7.3&(2.3)        \\
             & --0.070\,(0.013)         & --5.8\,(5.8) & 46.2&(14.1)       \\
\hline
\end{tabular}
\end{center}
a) From Gaussian fits adopting $z$=0.68466 and assuming that the continuum source covering factor is $f_{\rm c}$=1.
Standard deviations are given in parenthesis.
\end{table}

\begin{figure}[t]
\vspace{-0.3cm}
\centering
\resizebox{12.3cm}{!}{\rotatebox[origin=br]{-90}{\includegraphics{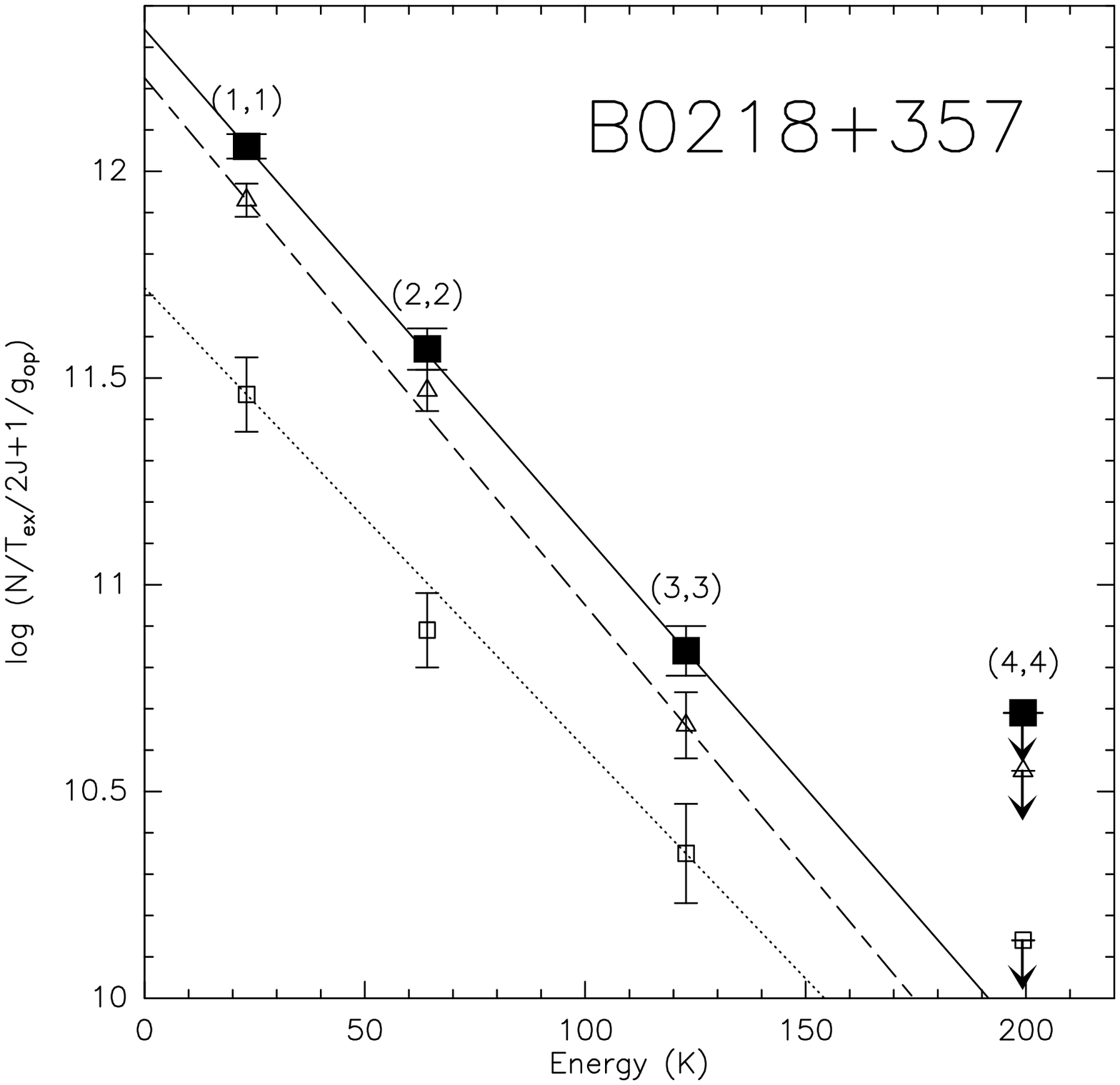}}}
\caption{Boltzmann plot (rotation diagram) of the normalized NH$_3$ column density ($g_{\rm op}$ is 1 for para-NH$_3$ ($K$ = 
1, 2, 4) and 2 for ortho-NH$_3$ ($K$ = 3)) divided by the excitation temperature $T_{\rm ex}$ in K within the inversion doublet 
as a function of excitation above the ground state. Lines connect the ($J,K$) = (1,1) and (3,3) normalized column densities 
(the (4,4) transition was not detected). Filled squares connected by a solid line: total NH$_3$ absorption; open triangles 
connected by a dashed line: broad NH$_3$ absorption component; open squares connected by a dotted line: narrow NH$_3$ absorption 
component. Assuming optically thin absorption and equal excitation temperatures in the inversion doublets, rotation temperatures 
become 35.5$\pm$2.5\.K (total), 34.0$\pm$2.9\,K (broad component) and 39.0$\pm$6.2\,K (narrow component).
\label{fig2}}
\end{figure}

\section{Molecular cloud properties}

\subsection{Location and linewidth}

Prior to this study, Menten \& Reid (\cite{menten96}) had searched for NH$_3$ absorption towards B0218+357. Their upper limits 
are higher than the strength of the absorption features shown in Fig.\,1. It is tempting to identify the two main features of 
the NH$_3$ lines as absorption against the two dominant continuum sources A and B. However, this is extremely unlikely
for several reasons. VLBI observations of H{\sc i} at dm- and of H$_2$CO at cm-wavelengths have shown that the absorption 
is confined to image A (Carilli et al. \cite{carilli00}; Menten \& Reid \cite{menten96}). While at radio wavelengths 
image A is three times as bright as B, image B is brighter at optical wavelengths. This indicates that image A is obscured 
by dust. Heavy absorption of A is further supported by a large differential rotation measure between A and B (Patnaik et al.
\cite{patnaik93}) and by the different separations between images A and B at radio and optical wavelengths (334$\pm$1 versus 
317$\pm$2\,mas, respectively; York et al. \cite{york05}). Apparently, radio and optical images of A do not coincide. We further 
note that images A and B are on opposite sides of the lensing galaxy. Assuming a rotation velocity of order $\ga$150\,km\,s$^{-1}$, 
a difference in radial velocity of $\sim$5\,km\,s$^{-1}$ between images A and B would then require an inclination $i$$\la$1$^{\circ}$ 
for the lensing galaxy. While the galaxy is clearly face-on (York et al. \cite{york05}), the optical image does not require 
an inclination that close to zero.

We conclude that the two velocity components are associated with image A that shows a core-jet morphology (Patnaik et al. 
\cite{patnaik95}; Porcas \& Patnaik \cite{porcas95}; Biggs et al. \cite{biggs03}). While the stronger narrower velocity 
component has a linewidth that is well within the range observed in galactic giant molecular clouds (e.g. Combes \cite{combes91}), 
the linewidth of the broad component is much larger. Could this be a consequence of hyperfine splitting of the ammonia lines? 
Or a consequence of differential galactic rotation? 

The NH$_3$ hyperfine satellites (for a ($J,K$)=(1,1) spectrum, see e.g. Ho \& Townes \cite{ho83}) are symmetrically bracketing 
the main feature and are covering $\sim$40\,km\,s$^{-1}$. Thus the partially blended satellite features from {\it both} velocity 
components can explain the observed linewidths of $\sim$50\,km\,s$^{-1}$. For a clear separation of individual satellite lines 
and velocity components, data with higher signal-to-noise ratios would be required. 

The jet extends from the core of A over $\sim$6\,mas toward the center of the lensing foreground galaxy, located at an angular distance 
of almost 300\,mas (Patnaik et al. \cite{patnaik95}; Porcas \& Patnaik \cite{porcas95}; Biggs et al. \cite{biggs03}; Wucknitz et al. 
\cite{wucknitz04}). Position angles are $\sim$68$^{\circ}$ for the jet (Biggs et al. \cite{biggs03}) and 65$^{\circ}$$\pm$2$^{\circ}$ 
for the center (Wucknitz et al. \cite{wucknitz04}). Therefore no significant velocity gradient is expected from the foreground galaxy. 
Its low inclination further reduces any such gradient. A wide component with an FWHP linewidth of order 50\,km\,s$^{-1}$ is not seen 
in other molecular lines at cm- and mm-wavelengths (e.g. Wiklind \& Combes \cite{wiklind95}; Menten \& Reid \cite{menten96}). Thus 
hyperfine splitting is likely the main cause for the linewidths observed. Note, that there are indications for the same asymmetry, 
i.e. a stronger feature at slightly positive and a weaker feature at slightly negative velocities, in the CO $J$=2--1 and 3--2, 
HCN 2--1 and HCO$^{+}$ 2--1 lines (Wiklind \& Combes \cite{combes95}; Combes \& Wiklind \cite{combes97}, \cite{combes98}). 

To summarize: While the relatively strong narrow absorption component at slightly positive velocities (see Fig.\,\ref{fig1}) 
must be optically thin, the optical depth of the wider and weaker component remains undetermined. Total linewidths are greatly
affected by the spacing of the hyperfine features. Gaussian fits to the wide component (Table 3) include the hyperfine 
satellites of both velocity components.

\subsection{Cloud size, optical depth and source covering factor}

Molecular lines at mm-wavelengths can show deep absorption (Combes \& Wiklind \cite{combes95}; Wiklind \& Combes 
\cite{wiklind95}; Combes \& Wiklind \cite{combes97}). Lines at dm- and cm-wavelengths are generally weaker with respect to the 
continuum. This even holds for $\lambda$=21\,cm H{\sc i} and 18\,cm OH, that show wider absorption than ammonia and the mm-wave 
lines (Carilli et al. \cite{carilli93}; Kanekar et al. \cite{kanekar03}) and that may thus arise from a larger volume. The peak line 
to total continuum ratios of the 18\,cm OH main lines are, however, similar to those seen in ammonia. 

Since the continuum sources are compact and since the measured NH$_3$ linewidths are small by extragalactic standards, the 
absorption arises from an area that must be considerably smaller than that studied in emission with single-dish telescopes towards 
nearby galaxies. According to Patnaik et al. (\cite{patnaik95}), $\sim$70\% of the total 22\,GHz flux is associated with image A. 
Thus minimum optical depths and source covering factors ($f_{\rm c,cm}$) of the ammonia lines toward image A are a factor of 1.4 higher 
than those suggested for the entire continuum flux by Fig.\,\ref{fig1}. They reach, in the (1,1) line, $\ga$0.014. 

Sensitive maps of image A show a source of size 10$\times$10\,mas$^{2}$ that is edge-brightened on its south-western side and 
tangentially stretched (Biggs et al. \cite{biggs03}). At 15\,GHz, the south-eastern region containing the core (components A1
and A2) and the north-western region containing the jet exhibit 62$\pm$1\% and 38$\pm$5\% of the observed flux density, respectively 
(Patnaik et al. \cite{patnaik95}). Although higher frequency VLBI measurements are missing, it would be no surprise if the 
contribution of the jet in image A is negligible at 100--150\,GHz, thus leaving only the lensed core as the background continuum 
source. Assuming for the core component a brightness temperature $T_{\rm b}$ and a flux density $S_{\nu}$ not drastically varying 
with frequency in the 10--150\,GHz frequency range (see e.g. Blandford \& K{\"o}nigl \cite{blandford79}), its solid angle would 
vary like $\Omega$ $\propto$ $\nu^{-2...-1}$ (see also Lobanov \cite{lobanov98}). According to NED\footnote{This research made 
use of the NASA/IPAC Extragalactic Database (NED) which is operated by the Jet Propulsion Laboratory, CalTech, under contract 
with NASA.} the mm-wave flux of B0218+357 is about half of that at cm-wavelengths, compatible with a fading jet and a flat 
spectrum core. 

Adopting this scenario and assuming that the absorption arises exclusively from the core as it is indicated by some of the 
deep mm-wave absorption lines (CO should arise from a larger volume than NH$_3$), we obtain for the peak of the ammonia (1,1) 
line a minimum optical depth and covering factor of 0.01/(0.7$\times$0.62) = 0.023. At 150\,GHz, this would yield for a cloud 
centered near the peaks of image A $f_{\rm c,mm}$ $\ga$ 0.15--1.0, depending on whether the exponent in the frequency dependence 
of the solid angle $\Omega$ is --1 or --2. 

High source covering factors at mm-wavelengths can be achieved, if the molecular absorber is elongated along the same position angle 
as the continuum, i.e. along a path with roughly constant galactocentric radius, hypothetically defining a spiral arm (see also
Sect.\,5). Elongated filaments are common in the Galaxy and are exemplified by the morphology of the Orion giant molecular cloud 
or the connection between this GMC and the Monoceros GMC (Maddalena \& Thaddeus \cite{maddalena85}), reaching linear sizes in 
excess of 100\,pc. This is more than the few 10\,pc that are needed here. Observed $f_{\rm c,mm}$ values near unity then suggest 
that actual source covering factors ($f_{\rm c,mm}$ and $f_{\rm c,cm}$) are 1--6 times higher than the lower limits estimated above, 
not providing an additional constraint for the NH$_3$ source covering factor.

To verify these estimates and to further elucidate the morphology of image A and its foreground molecular cloud at high frequencies, 
mm-wave VLBI observations are highly desirable.

\subsection{Densities and column densities}

So far, the nature of the molecular cloud has been an enigma. It was proposed to be either a diffuse or a dark cloud (Combes \& Wiklind 
\cite{combes95}; Wiklind \& Combes \cite{wiklind95}; Menten \& Reid \cite{menten96}; Kanekar et al. \cite{kanekar03}). The determination 
of an NH$_3$ rotation temperature, which is a lower limit to the kinetic temperature, allows us to discriminate between these two 
scenarios. Provided total gas densities are less than $n$(H$_2$)$\sim$10$^{6}$\,cm$^{-3}$ and using the Large Velocity Gradient 
(LVG) model of Schilke (\cite{schilke89}) with collision rates from Danby et al. (\cite{danby88}), the kinetic temperature is 50--60\,K. 
For even higher densities the kinetic temperature would be lower but would still be $>$35\,K. Since frequencies of the ammonia lines 
are quite similar, the resulting temperatures are not significantly affected by potential differences in the morphology of the 
background continuum. We therefore conclude that the NH$_3$ absorption {\it does not arise from a cool dark cloud}. Since the 
HCO$^+$ absorption lines are not fully saturated (Wiklind \& Combes \cite{wiklind95}) and many molecules have not been detected 
at levels of $<$1\% of the total continuum flux density (Table 2), {\it the diffuse cloud scenario is preferrable}.

The high kinetic temperature of the gas implies that our non-detection of the DCN $J$=1--0 line does not provide significant 
constraints to the cosmic D/H ratio. Adopting a width of the line of order 5\,km\,s$^{-1}$, we deduce a 5$\sigma$ sensitivity 
of 5\% of the continuum level (Table 2). Assuming an optical depth of 5 for the opacity of the HCN $J$=1--0 main line, this 
implies that we would have seen a DCN line only if [DCN]/[HCN]$\ga$10$^{-2}$. A ratio of 10$^{-2}$ is, however, not reached at 
$T_{\rm kin}$=20\,K (Gerin et al. \cite{gerin99}) and [DCN]/[HCN] ratios at higher temperatures should be even smaller.

Having estimated the kinetic temperature, we can use measurements of molecular excitation to also obtain the density of the gas 
toward B0218+357. From the microwave background we obtain a minimum of $T_{\rm cmb}$ = 2.73 (1+$z$) = 4.6\,K. Combes et al. 
(\cite{combesetal97}) find $T_{\rm ex}$ = 6$^{+3}_{-2}$\,K from CS $J$=1--0 and 3--2. The detection of the 557\,GHz and the 
non-detection of the 183\,GHz line of water vapor (these are rest frequencies) indicate $T_{\rm ex}$ $<$30\,K (Combes \& Wiklind
\cite{combes97}). $^{13}$CO $J$=2--1 and 4--3 spectra suggest $T_{\rm ex}$$<$20\,K (Gerin et al. \cite{gerin97}). With 
$T_{\rm kin}$ $\ga$ 35\,K and $n$(H$_2$) $\ga$ 10$^{6}$\,cm$^{-3}$, LVG calculations for CS (for details, see e.g. Wang et al. 
\cite{wang04}) yield an excitation far above that observed. Thus $T_{\rm kin}$ $\sim$ 55\,K. With this kinetic temperature and 
$T_{\rm ex,CS}$$\leq$9\,K, $n$(H$_2$)$\la$2$\times$10$^{4}$\,cm$^{-3}$. This is a firm upper limit. If the excitation is close 
to 6\,K or if the stronger CS line ($J$=1--0) is optically thick, densities are smaller ($\sim$7$\times$10$^{3}$\,cm$^{-3}$ in 
the former case), while constraints from H$_2$O are less stringent. Again using LVG calculations, we find that a density of 
$\sim$5$\times$10$^{3}$\,cm$^{-2}$ is consistent with $^{13}$CO excitation temperatures $\la$20\,K, if optical depths are smaller 
than unity. Since Combes \& Wiklind (\cite{combes95}) find that the $^{13}$CO $J$=2--1 line is saturated, however, the 
density will be even lower. The upper density limits derived here are also consistent with the upper limit to the NH$_3$ (2,1) 
line (Fig.\,\ref{fig1} and Table 2), that would only be detectable in a cloud with much higher density (see Mauersberger et al. 
\cite{mauers85}). 

Adopting $T_{\rm kin}$=55\,K and, less certain, $n$(H$_2$)=5$\times$10$^{3}$\,cm$^{-3}$ as best guesses, we can also 
estimate the excitation temperature within an NH$_3$ inversion doublet: $T_{\rm ex}$$\sim$6\,K (Schilke 1989, his Fig.\,6.15). 
This can be used to derive total NH$_3$ column densities. Adding up column densities in the three detected lines gives $N$(NH$_3$) 
= 4$\times$10$^{13}$\,cm$^{-2}$. This is a minimum value. With equation A15 of Ungerechts et al. (\cite{ungerechts86}) and rotation 
temperatures of $T_{12}$=35--40\,K, we obtain $\sim$5$\times$10$^{13}$\,cm$^{-2}$. The error lies mainly in the assumed density of 
the gas. $n$(H$_2$)=10$^{4}$\,cm$^{-3}$ implies an excitation temperature of 8\,K and a column density of 5 and 
7$\times$10$^{13}$\,cm$^{-2}$, respectively. So far all given column densities are averaged over the entire continuum source. 
Averaged over the core of image A, column densities have to be multiplied by a factor of 2.3.

Typical galactic clouds have fractional abundances of $X$(NH$_3$) = 10$^{-8}$--10$^{-7}$ (e.g. Hotzel et al. \cite{hotzel04}).  
Averaged over the entire continuum source, this yields $N$(H$_2$) $\sim$ 4$\times$10$^{20}$ -- 7$\times$10$^{21}$\,cm$^{-2}$  
or 1$\times$10$^{21}$ -- 2$\times$10$^{22}$\,cm$^{-2}$ for the core of image A. The lower limit is too small to allow us to detect 
a variety of molecular lines. Therefore the fractional abundance is likely close to 10$^{-8}$. Our column density range is 
slightly lower than what has been proposed by Menten \& Reid (\cite{menten96}) on the basis of another molecular cm-wave 
transition. Adopting 55\,K also for the H{\sc i} spin temperature, the total background continuum averaged H{\sc i} column density 
becomes 2$\times$10$^{20}$\,cm$^{-2}$ (see Carilli et al. \cite{carilli93}). Accounting for an image A core source covering 
factor, the column density becomes 5$\times$10$^{20}$\,cm$^{-2}$. Thus the line of sight must be mostly molecular.

Column densities averaged over the extent of the continuum source are lower at cm-wavelengths (see Menten \& Reid \cite{menten96})
then those derived by the higher frequency studies of Gerin et al. (\cite{gerin97}), $\ga$2$\times$10$^{22}$\,cm$^{-2}$, and 
Combes \& Wiklind (\cite{combes95}), $\sim$5$\times$10$^{23}$\,cm$^{-2}$. This is expected for a molecular cloud located near the 
peak of image A and covering most of the mm-wave but only a fraction of the cm-wave continuum of B0218+357. In spite of this effect 
the value proposed by Combes \& Wiklind (\cite{combes95}), $N$(H$_2$)$\sim$5$\times$10$^{23}$\,cm$^{-2}$, must be too large. With 
such an enormous column density many of the species listed in Table 2 would have been detected. The $^{18}$O/$^{17}$O ratio of 
$\sim$15 suggested by Combes \& Wiklind (\cite{combes95}), that is in part responsible for the high column density, is far outside the 
range observed so far in interstellar clouds. With $^{18}$O/$^{17}$O$\sim$1.6 in the LMC (Heikkil{\"a} et al. \cite{heikkilae98}), 
$\sim$4.1 in the local interstellar medium (Wouterloot et al. \cite{wouterloot05}), 5.5 in the solar system (e.g. Wilson \& Rood 
\cite{wilson94}), and $\sim$6.5 in nearby starburst galaxies (Harrison et al. \cite{harrison99}; Wang et al. \cite{wang04}), new 
measurements of rare CO isotopomers would thus be desirable.

\section{A comparison with `local' molecular clouds}

To put our data into context with the much more extensively studied interstellar medium of the local universe is not easy. The 
kinetic temperature we derive, $\sim$55 K, is much higher than the canonical number of 10\,K assumed for quiescent Giant 
Molecular Cloud gas in the Galaxy. Clouds forming massive stars are warmer and may reach 55\,K, but densities tend to be larger 
than those estimated in Sect.\,4.3 for B0218+357. 

The central region of the irregular starburst galaxy  M~82 may provide the best agreement with the molecular parameters of 
B0218+357. In M~82, NH$_3$ observations indicate $T_{\rm rot}$$\sim$30\,K and $T_{\rm kin}$$\sim$50\,K (Wei{\ss} et al. 
\cite{weiss01}), while observations of CO reveal densities of several 10$^{3}$\,cm$^{-3}$ (e.g. Mao et al. \cite{mao00}). 
The interstellar medium of M~82 is characerized by widespread PDRs (photon dominated regions/photodissociation regions), 
where ammonia, being readily destroyed by UV photons, mostly resides in the few remaining well shielded relatively cool 
cloud cores. Fractional ammonia abundances are of order 5$\times$10$^{-10}$ and thus much lower than in B0218+357. 
In the inner few 100\,pc of nearby spiral galaxies, detected in NH$_3$, temperatures are significantly higher 
than in B0218+357 ($T_{\rm rot}$$>$35\,K; $T_{\rm kin}$$>$55\,K; Mauersberger et al. \cite{mauers03}). This also 
holds for clouds within a few 100\,pc of the center of our Galaxy (G{\"u}sten et al. \cite{guesten85}; Mauersberger et al. 
\cite{mauers86}; H{\"u}ttemeister et al. \cite{huette93}).

In the Galaxy, few measurements of NH$_3$ absorption from diffuse clouds against radio continuum sources have been reported and 
neither this molecule's abundance (if detected) nor its rotation temperature are known in such environments (Nash \cite{nash90}). 
It would thus be easiest to reconcile our NH$_3$ results with a known scenario if we assumed that the B0218+357 absorption line
of sight crossed the central region of the lensing galaxy. This, however, is not the case. Wucknitz et al. (2004) model the 
B0218+357 system using the L{\small ENS}C{\small LEAN} algorithm and find the lens to be centered at a position which is 
($\Delta \alpha$, $\Delta \delta$) = (255, 120)\,mas offset from the A image and (--55,--10)\,mas from the B image, with 
uncertainties of order 5\,mas in each coordinate. The best fit size of the lens is 258$\times$121\,mas with a position angle of 
PA=--48$^{\circ}$ east of north (Wucknitz et al. \cite{wucknitz04}, Table 4). This implies that image A is at a projected distance 
of $\sim$2.0\,kpc (P.A.=--115$^{\circ}$) from the center of the lensing mass distribution, while image B is offset by only 0.4\,kpc 
(PA=+70$^{\circ}$). Since the starburst in M~82 is, as most starbursts, confined to the nuclear few 100\,pc and since its 
fractional NH$_3$ abundance is much smaller than that in B0218+357, we do not find, neither in the Galaxy nor in the nearby 
extragalactic universe, any analog to the line-of-sight toward image A in B0218+357.

\section{Conclusions}

The gravitational lens system B0218+357 allows us to study the details of a molecular cloud observed at about half a Hubble 
time in the past. Our analysis of the Effelsberg data reveals the following:

\begin{itemize}

\item Ammonia (NH$_3$) is detected in absorption in its three lowest metatstable inversion doublets, while the ($J,K$)=(4,4) 
and the non-metastable (2,1) lines remain undetected. Upper limits are also determined for transitions of SO, DCN, OCS, SiO, C$_3$N, 
H$_2$CO, SiC$_2$, HC$_3$N, HC$_5$N, and CH$_3$OH.

\item NH$_3$ is likely absorbing exclusively the core of image A. Source covering factors and peak optical depths are
$\ga$0.023 toward this core. 

\item The NH$_3$ rotation temperature of $T_{\rm rot}$ $\sim$ 35\,K indicates a kinetic temperature of $T_{\rm kin}$ 
$\sim$ 55\,K for the molecular gas. 

\item With the kinetic temperature obtained from NH$_3$, the measured excitation of other molecular species can be used to 
constrain the density. $n$(H$_2$) $\sim$ 5$\times$10$^3$ is found to be a plausible value. Averaged over the entire continuum
source, the NH$_3$ column density is $N$(NH$_3$) $\sim$ 5$\times$10$^{13}$\,cm$^{-2}$; averaged over the core of image A, 
it is $\sim$1$\times$10$^{14}$\,cm$^{-2}$ with a fractional abundance of order 10$^{-8}$ . The gas along the line of sight is 
predominantly molecular. 

\item The physical parameters are quite peculiar, when compared with those of clouds in the local universe.

\item Because the morphology of the radio continuum depends on frequency, cm-wave and mm-wave absorption features are characterized 
by different H$_2$ column densities. 

\end{itemize}

\begin{acknowledgements}
We wish to thank R. Porcas, C. B{\"o}ttner, M. Kadler and T. Krichbaum for useful discussions and an anonymous referee for critically 
reading the text and making a number of important suggestions.
\end{acknowledgements}

\end{document}